\documentclass[jpcm]{iopart}
\usepackage{epsf,citesort}
\usepackage{graphicx,citesort,amssymb}

\begin{document}

\title{Dilatancy, Jamming, and the Physics of Granulation}

\author{M. E. Cates, M. D. Haw, and C. B. Holmes}
\address{School of Physics, JCMB Kings Buildings, The University of
Edinburgh, Mayfield Road, Edinburgh EH9 3JZ, United Kingdom}

\begin{abstract}
Granulation is a process whereby a dense colloidal suspension is converted into pasty granules (surrounded by air) by application of shear. Central to the stability of the granules is the capillary force arising from the interfacial tension between solvent and air. This force appears capable of maintaining a granule in a jammed solid state, under conditions where the same amount of solvent and colloid could also exist as a flowable droplet. We argue that in the early stages of granulation the physics of dilatancy, which requires that a powder expand on shearing, is converted by capillary forces into the physics of arrest. Using a schematic model of colloidal arrest under stress, we speculate upon various jamming and granulation scenarios. Some preliminary experimental results on
aspects of granulation in hard-sphere colloidal suspensions are also presented.

\end{abstract}

\maketitle

\section{Introduction}
There is obviously a connection between colloid physics and granular matter, but the relationship is subtle when looked at in detail. Colloidal suspensions do not suffer static interparticle friction and do have significant Brownian motion; this should make them much more flowable than dry granular media. On the other hand, colloids occupy a fixed volume of incompressible fluid: so long as they remain in a homogeneous bulk phase, they cannot expand their volume under shear. Since dense powders must dilate before they can flow \cite{reynolds}, colloidal dynamics is impeded by this fixed-volume constraint.

It is known that hard-sphere colloids develop a yield stress at the glass transition, empirically found at volume fraction $\phi = \phi_g \simeq 0.58$. This number is well below random close packing ($\phi_{rcp} = 0.64$ \cite{rcp,rcp2}) but suggestively close to the less well defined ``random loose packing'' limit which is, roughly speaking, the lowest density of dry grains capable of sustaining solid-like behaviour \cite{loosepack}. Moreover $\phi_g$ is also quite close to the critical state volume fraction $\phi_c$ above which dry grains must dilate in order to flow \cite{criticalstate,loosepack}. However, no direct link is yet known between the hard sphere glass transition and either the random loose packing or the critical state, although these two may well be related to one another \cite{loosepack}.

An interesting phenomenon which combines elements of colloidal and granular behaviour is that of ``granulation'' \cite{granulationrefs,grannote}. In this process, a very dense colloidal suspension is subject to sustained shearing. It is found that the sheared sample first jams, and then fractures into lumps (with ingress of air, creating large amounts of air-solvent interface). These lumps rub against each other, break, perhaps coalesce, and generally form a complicated mess, until eventually the system settles down into a state with a relatively uniform and reproducible size distribution for the lumps, by now dignified with the name ``granules''. This process is used industrially in the manufacture of products ranging from pharmaceutical preparations to washing powder \cite{granulationrefs}. A typical granule size might be tens to hundreds of microns, containing between hundreds, and tens of thousands, of colloidal particles. Theoretical models of this process have been widely developed in recent years \cite{granulationmodels} but these mainly address the evolution of the granule size distribution using nucleation and growth and/or fragmentation and coalescence ideas, often with a sophisticated dependence on parameters such as viscosity, contact angles, etc.. In this paper we ignore this aspect, despite its obvious technological importance. We address instead the fundamental and neglected question of what is happening in the earliest stages of granulation when a homogeneous suspension initially breaks up. We also address the physics of the granules themselves, which turns out to be an intimately related question.

In fact, the granules created by the granulation process just described have fascinating properties whose physical exploration is frustratingly incomplete in the literature. We summarise here what can be gleaned from side remarks contained in papers primarily addressing other issues \cite{bibette,haw}, and by private communication \cite{warren,bibettepc}. Some of these phenomena are reported in Ref. \cite{bibette} for a low-density gelling colloid in which strong attractive interactions are surely present, but below we report
somewhat similar results for well-controlled hard-sphere colloids at concentrations close to their glass transition. Similar observations have been made in suspensions of zeolite in amphiphilic solvents \cite{warren}. It is too soon to say whether or not these features are ubiquitous in all granulating suspensions. The observed phenomenology is, broadly speaking, as follows.

First, granules have a matt appearance to the eye, and under microscopy show irregularity of surface shape and/or particles protruding through the interface. Second, they hold their irregular shapes indefinitely. Third, granules are bistable: if a granule is placed on a plate which is then vibrated, it melts from its irregular shape to a spherical one. At the same time, the matt surface becomes glossy, showing that particles no longer protrude significantly through it. (Colloidal particles have radii of order the wavelength of light, so protrusion gives strong scattering.) The resulting object is, in fact, no longer a solid granule but a flowable droplet. A similar and equally striking experiment involves placing such a melted droplet in contact with a granule that is still frozen. In this case, the frozen granule is rapidly coalesced and the result is a single, melted droplet of larger size \cite{warren}. Finally, a flowable spherical droplet can sometimes be converted back into a mis-shapen granule with a matt surface, simply by prodding it firmly with a spatula.

It is not yet clear whether the flowable droplet state is always strictly fluid, in the sense of having finite zero shear-viscosity; this would require the conditions within the interior of the droplet to be within the colloidal fluid phase. In some cases, the flowable droplet may instead be in the glass phase (see Sections \ref{expt}, \ref{rheology} below), but have a small enough yield stress that it behaves as a fluid under the prevailing experimental conditions. In any case, it is much more flowable than the granule it was derived from.

These results shows that the same amount of colloid and solvent can exist in two quite different states, one solid, one liquefied. The solidity of the granule forces the solvent to adopt a rough surface, creating capillary stresses that can be extremely large: anything up to about $\Sigma/a$ is possible, with $a$ the radius of a colloidal particle and $\Sigma$ the interfacial tension. Moreover, so long as it remains solid, these capillary forces can translate into off-diagonal (shear) stresses within the bulk of the granule. The liquid state of the same granule is a spherical droplet with Laplace pressure $2\Sigma/R$, where $R$ is the droplet radius; so long as the interior is indeed a liquid, this pressure (even if large) remains isotropic and no shear stresses can develop within. 

In our view, the observed bistability presents compelling evidence for the jamming of a colloidal suspension under static shear stresses \cite{jamming}. These stresses are generated in a self-consistent manner; they originate in the capillary forces at the surface of the granule, but are transmitted internally via the jammed solid itself. However, externally applied deformation is seemingly required to set up the jammed solid initially.

Note that, because the volume of solvent is fixed, the protrusion of particles through the surface of a granule means that the volume fraction $\phi$ of colloid within its interior solvent is marginally {\em below} that of the corresponding droplet. It follows that the jammed solid within a granule is slightly dilated. 

\section{Experimental results}
\label{expt}
The above summary of the phenomenology is based in part on our own work on well characterised hard-sphere suspensions. In this Section we present some of this preliminary work; a more comprehensive study will appear elsewhere \cite{hawpc}. Here we consider volume fractions lying above the glass transition so that the system at rest has a small yield stress (see comments above).
 
\begin{figure}
\begin{center}
 \epsfxsize=8cm
   \leavevmode\epsffile{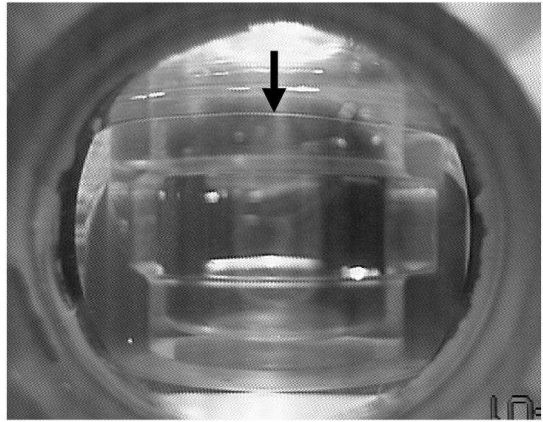}
	\caption{Refractive-index matched PMMA hard-sphere suspension, particle radius $\approx 270$ nm, at volume fraction $\Phi\approx 0.63$, sheared in a transparent cylindrical couette cell, gap width 0.5mm, shear rate $\gamma \approx 17 $s$^{-1}$.  The arrow shows the top surface of the suspension, above which the gap is filled with air.  At this shear rate the flow is regular.}
\label{fig1_couettelowshear}
\end{center}
\end{figure}

\begin{figure}
\begin{center}
 \epsfxsize=8cm
   \leavevmode\epsffile{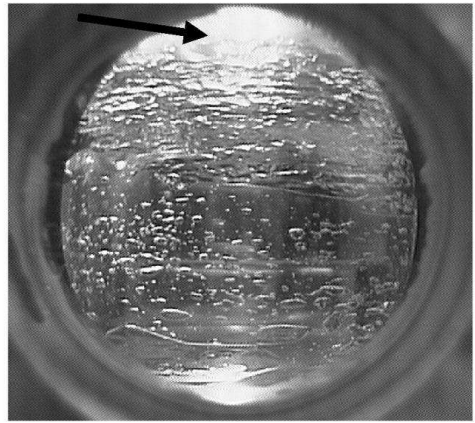}
	\caption{PMMA hard-sphere suspension at volume fraction $\Phi \approx 0.63$, 
sheared as in Figure \protect{\ref{fig1_couettelowshear}} but at shear rate $\gamma \approx 50 $s$^{-1}$.  The arrow shows the top surface of the suspension.  Shearing at high rate leads to the inclusion of air,
while parts of the suspension appear almost solid, resisting strain.}
\label{fig2_couettehighshear}
\end{center}
\end{figure}

Figures \ref{fig1_couettelowshear} and \ref{fig2_couettehighshear} show how the flow of a dense colloidal suspension becomes irregular and leads to ingress of air if the flow rate is high enough. This is a sample of $\approx 270$ nm radius sterically stabilized polymethylmethacrylate (PMMA) colloids, refractive index matched (ensuring effectively hard-sphere interactions) sheared in a transparent couette cell.  The behaviour
is reversible, in the sense that returning the shear to lower rate leads to (slow) coalescence of air bubbles
and eventual expulsion of the air, the sample regaining a regular surface.  
However, if the high shear rate run is stopped abruptly, the sample remains frozen (for hours at least) with air-bubbles still in place.

\begin{figure}
\begin{center}
 \epsfxsize=8cm
   \leavevmode\epsffile{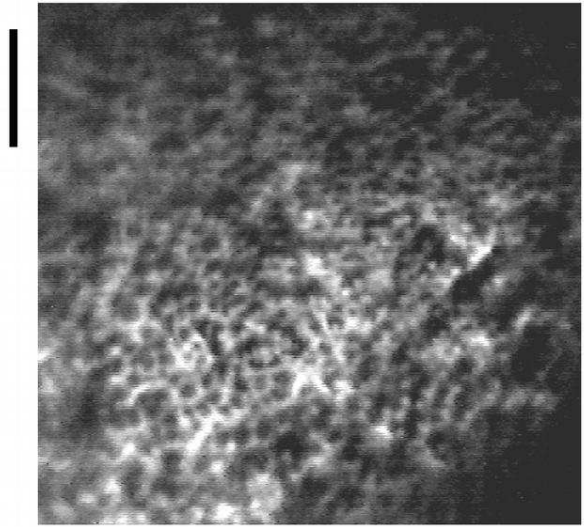}
	\caption{Microscope image of the suspension-air interface at the edge of a PMMA hard-sphere suspension, particle radius $\approx 1000$ nm, at volume fraction $\Phi \approx 0.60$.  The suspension has been squeezed between approximately parallel microscope slides. The scale bar at top left is 20$\mu$m.}
\label{fig3_squeeze}
\end{center}
\end{figure}

Figure \ref{fig3_squeeze} shows a microscopy image of the interface with air of a confined `granule' of a dense colloidal suspension containing $1\mu$m radius particles. This was created by taking a fluid droplet of the suspension and squeezing it between two parallel microscope slides. The overall volume fraction of colloid does not change during this process, yet the image shows particles protruding through the air/solvent surface (creating much higher refractive index contrast than present in the bulk of the sample). A fully fluid sample in the same geometry would present a featureless, smooth surface (not shown).

\begin{figure}
\begin{center}
 \epsfxsize=8cm
   \leavevmode\epsffile{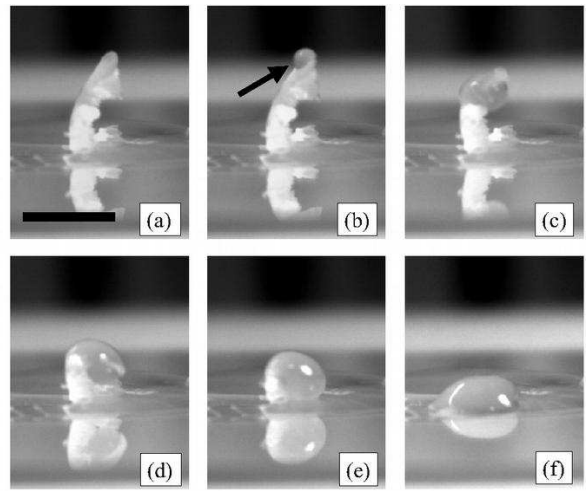}
	\caption{A sample of refractive-index matched PMMA hard-sphere suspension, particle radius
 $\approx 1000$ nm, at $\Phi \approx 0.61$, partly `granulated' by pushing a spatula through a droplet.  In (a) the granulated, opaque, solid piece of the sample can be seen; this persists indefinitely if left undisturbed.  (b) to (f) show melting of the granulated
sample by contact with a droplet of a more dilute solution ($\Phi \approx 0.3$) of the same particles (the arrow
in (b) shows the initially deposited fluid droplet).  The melting takes approximately one
second. The scale bar in (a) is 5mm.}
\label{fig4_granandcontactmelt}
\end{center}
\end{figure}

\begin{figure}
\begin{center}
 \epsfxsize=8cm
   \leavevmode\epsffile{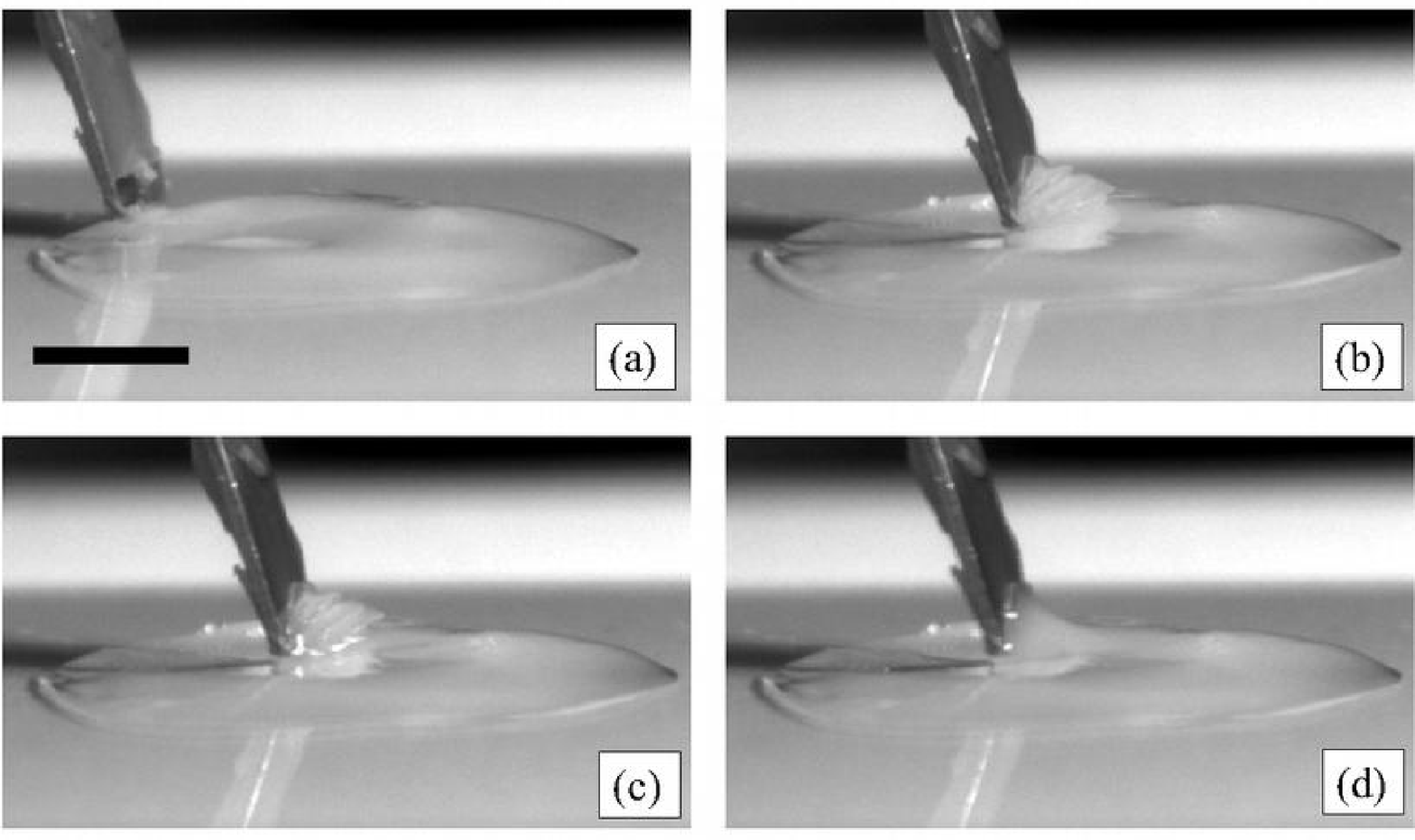}
	\caption{A droplet of refractive-index matched PMMA hard-sphere suspension, particle radius
 $\approx 1000$ nm, at $\Phi \approx 0.61$, showing granulation caused by pushing a spatula through the droplet.  The frames are taken approximately every 0.25 seconds.  In (a) and (b) the 
creation of an opaque, rough-surfaced region due to the stress applied by the spatula can be seen.  On cessation of pushing with the spatula [(e) and (f)], after a delay of about 0.5 seconds the granulated opaque region melts, its surface becoming glossy and liquidlike again. This does not happen unless the granulated region is still in contact with non-granulated fluid. The scale bar in (a) is 5mm.}
\label{fig5_granbypushing}
\end{center}
\end{figure}

Figure \ref{fig4_granandcontactmelt} shows a granular lump [(a)] created simply by 
pushing a spatula by hand through a larger drop. The solid-like nature of the granular lump is apparent. Then (b) to (f) show the melting of the granule by contact with a fluid droplet. In this case, the added fluid droplet contains significantly lower colloid concentration than the granular lump. Figure \ref{fig5_granbypushing} shows a similar effect, again involving formation of a granular lump by the same process of pushing a spatula. Now, however, the lump remains in contact with the larger droplet from which it was created. If the granule were
pushed all the way out of the droplet it would remain frozen even on removal of
stress by stopping pushing (as shown in (a) of 
Fig.~\ref{fig4_granandcontactmelt}).
But now, when the granule comes to rest, it remelts. This can be interepreted
as melting of a granule by contact with a fluid droplet of similar volume fraction (albeit the one from which it was drawn). These preliminary results are slightly ambigious since we cannot rule out a slight increase in the colloid density during the collection of the initial granule by a self-filtration effect as reported in \cite{haw}. Nonetheless they are consistent with reports of the melting of
a granule by contact with a fluid droplet, even when both are drawn from a fluid of the same colloid density \cite{warren}. We hope to confirm or disprove that phenomenon in hard sphere suspensions (rather than zeolite \cite{warren} or flocculated colloids \cite{bibette}) in future work \cite{hawpc}.

\section{Colloid rheology}
\label{rheology}
In contrast to dry powders, colloidal glasses with volume fractions $\phi$ just above $\phi_g$ are generally flowable without dilatancy -- indeed these materials are often drastically shear thinning at low shear rates \cite{brady}. This difference can perhaps be attributed to Brownian motion. Yet the same materials at higher shear rates and concentrations can show drastic shear thickening. Moreover, materials just below the glass transition can also shear-thicken, and in some cases an erratic flow regime is observed \cite{benderwagner,laun,erraticprl}, suggestive of a jamming transition. (This happens even in samples that are apparently fluids, not glasses, when at rest.) Although hydrodynamic theories of shear thickening have been developed \cite{brady}, there is strong evidence that without Brownian motion the flow of hard spheres is singular and leads to complete arrest in finite time \cite{melrose}; details of the interaction potentials then dominate.
In any case, attempts to model nonlinear colloid rheology hydrodynamically without accounting for the underlying glass transition are lacking an essential physical ingredient \cite{epl,bbk,fuchs}. For example they predict a finite zero-shear viscosity for $\phi_g<\phi<\phi_{rcp}$, in contrast to observation \cite{poonvisco}. 

To start to fill this gap in our theoretical understanding of colloid rheology, we developed in Refs.\cite{epl,jrheol}
an approach to shear thickening based on the glass transition alone, which largely ignores hydrodynamics. (Obviously it would be good to incorporate both elements at some point; but this is a future challenge.) Our approach builds on a mode coupling theory (MCT) for colloids under shear \cite{fuchs}; for related work see \cite{reichman}. The MCT of \cite{fuchs} predicts purely shear-thinning behaviour and no shear thickening. This lack of a shear-thickening regime might result from shortcomings in some standard approximations implicit in MCT, which do not distinguish between hard and soft particle interactions; the only information that enters is the static structure factor $S(q)$ for the quiescent state \cite{paris}. Shear thickening and jamming could well involve many-body correlations, for example via the formation of load-bearing force-chains extending across many particle radii \cite{jamming}; these cannot be picked up in the static $S(q)$. This deficit is rectified, in an ad-hoc but interesting way, in Refs. \cite{epl,jrheol} where we introduce an explicit stress dependence to the MCT vertex. This is done within a much simplified ``schematic'' model of the glass transition, which nonetheless captures, in cartoon form, the more elaborate calculations of \cite{fuchs}. 

\section{Phase diagram of the schematic jamming model}
\label{colin}

The basic physics represented by this schematic model is a contest between the effects on particle organisation of the strain rate $\dot\gamma$ and the shear stress $\sigma$. Strain reorganises local environments, and abolishes the memory effects that result from colloidal caging (which MCT is intended to capture). In contrast to this, we argue that stress (which for simplicity is treated in a scalarised fashion \cite{leshouches}) promotes arrest, by jamming particles into contact. Our schematic model involves two free parameters, a ``glassiness'' parameter $v_2$, and a ``jammability'' $\alpha$. More precisely, $v_2$ is the static MCT vertex amplitude and controls the quiescent glass transition (for $v_2<v_c = 4$ the system is fluid, for $v_2>4$ it is a glass; the critical value of $4$ is inherited from earlier schematic models of the quiescent state \cite{schematics}). The jammability parameter $\alpha$ determines the strength of the dependence of the MCT vertex on stress. The model comes in several variants, whose details need not concern us here \cite{jrheol}. 

\begin{figure}
\begin{flushright}
\includegraphics[width=0.8\textwidth]{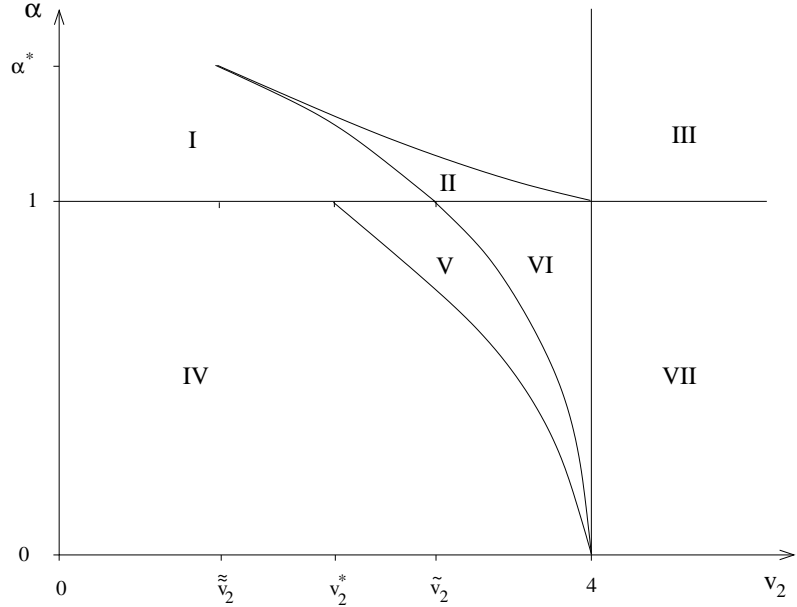}
\end{flushright}
\caption{Phase diagram on the $(v_2,\alpha)$ plane for one variant of the schematic model of \cite{epl,jrheol} as described in the text. The marked points on the axes have the following values: $\alpha^* = 27-15\sqrt{3}$; $\tilde{\tilde v}_2 = 9(1-1/\sqrt{3})$; $v_2^* = 3.815\pm 0.005$; $\tilde{v}_2 = 1+2\sqrt{2}$. These are specific to the model variant, as is the detailed behaviour around the point $(v_2,\alpha) = (4,1)$, and the absence, in this variant of the model, of regimes Ia, IIa, VIa as defined in the text.} \label{phasediag}

\end{figure}

In Figure \ref{phasediag} we present the phase diagram of the model in the $(v_2,\alpha)$ plane for one of these variants (in the terminology of \cite{jrheol}, it is model I). Before discussing the various phases, note that both the glassiness $v_2$ and the jammability $\alpha$ can depend on all aspects of particle interactions --- even though the schematic approach offers no theory of what this dependence should be. Therefore, as volume fraction $\phi$ is varied, some trajectory on the $(v_2,\alpha)$ plane is traced out, but this trajectory is interaction-dependent and can vary from one colloidal system to another. This admits a wide range of scenarios for evolution with concentration of the steady-state flow curve $\sigma(\dot\gamma)$, some of which we discuss below.
 
Each of regimes I--VI in Figure \ref{phasediag} corresponds to a qualitatively distinct flow curve $\sigma(\dot\gamma)$ for a system undergoing simple shear, as enumerated in Section \ref{regimes} below. Note that other, more elaborate curves could result from different variants of the model \cite{jrheol}. Specifically, there are re-entrant flow curves in regimes I, II and VI, all of which start off with a Newtonian section (no yield stress) at small stresses. In a more general picture each of these could extend beyond the static glass line (here, $v_2=v_c=4$) to give a similar flow curve, but with an additional yield threshold at the lowest stresses. In the granulation context this is linked to an issue raised above, of whether a droplet made by liquefying a granule has a yield stress or not. By default we will assume not, but can allow for one by expanding our model space slightly and invoking the corresponding additional regimes which we will call Ia, IIa, and VIa.

\subsection{The various regimes}
\label{regimes}

We now enumerate the regimes in Figure \ref{phasediag}. These are implicit in the work of \cite{jrheol}, although our presentation here in the form of a phase diagram is new.

\subsubsection{Ultimately yielding regimes:}

\begin{figure}
\begin{flushright}
\includegraphics[width=0.8\textwidth]{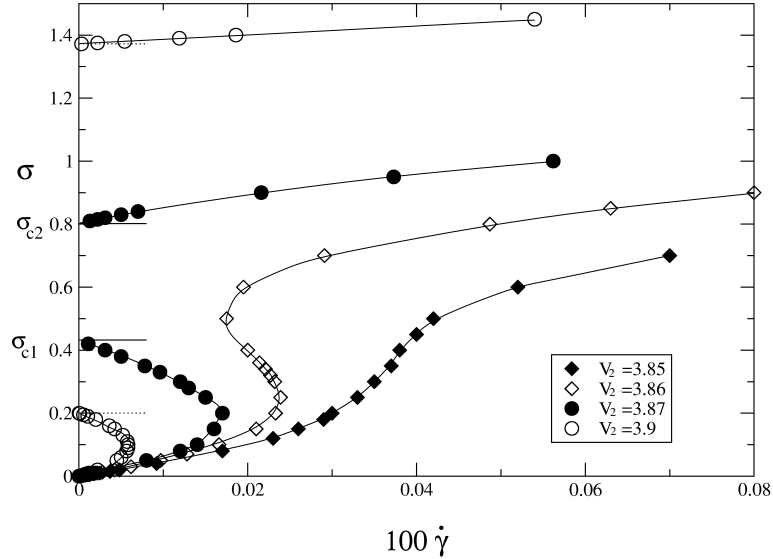}
\end{flushright}
\caption{Flow curves for regimes IV (continuous shear thickening, solid diamonds), V (discontinous shear thickening, open diamonds), and VI (full jamming, open and closed circles). The units on both axes are set by the schematic model and arbitrary for present purposes (but see \protect{\cite{jrheol}} for a discussion of stress and time scales).} \label{fulljam}

\end{figure}

We start with the sequence IV--VII which arises on increasing $v_2$ for $\alpha < 1$; in this case, all curves have an ultimately shear-thinning behaviour at very high stresses. Put differently, for $\alpha < 1$ the jammability is not sufficient to prevent the ultimate yielding of the material, through a homogeneous flow mechanism, at sufficiently large $\sigma$. The regimes are:

IV: A monotonic flow curve, with either shear thinning or shear thickening (or in some cases both). A shear-thickening example is shown in Figure \ref{fulljam}.

V: A re-entrant, S-shaped flow curve (also shown in Figure \ref{fulljam}). This admits discontinuous shear thickening via a shear-banding mechanism in which the material creates shear bands (with layer normals in the vorticity direction). At the discontinuous thickening transition the system jumps from the lower to upper branch \cite{epl,jrheol}. Note that any decreasing sections of the flow curve are mechanically unstable and, as a rule of thumb, are always bypassed by shear banding \cite{banding} or a similar process.

VI: A ``full jamming'' flow curve. The upper left branch of the S-shaped curve has now collided with the vertical axis, giving a vertical segment of the flow curve at zero shear rate $\dot\gamma$ but finite stress $\sigma_{c1} < \sigma < \sigma_{c2}$ (Two examples are shown in Figure \ref{fulljam}). Within this window the material is fully arrested \cite{head}, with zero steady flow rate so long as the stress level is maintained. (Note that creep flow, if sublinear in its time dependence, is not excluded.) This regime describes a system that is fluid at rest, but jams on increasing stress and then unjams again beyond $\sigma_{c2}$. 

VII: A yielding colloidal glass. The flow curve has a yield stress $\sigma_y$ with zero $\dot\gamma$ for $\sigma<\sigma_y$ and shear thinning (downward curvature) beyond. This is the flow curve predicted for glasses within standard MCT \cite{fuchs} as recovered here in the limit $\alpha = 0$. It is obtained from the full jamming curve by setting $\sigma_{c1} = 0$ and $\sigma_{c2} = \sigma_y$.

\subsubsection{Ultimately jamming regimes:}

We now turn to the regimes I--III arising for $\alpha >1$. (As visible in Figure \ref{phasediag}, these can occur in the sequence I,II,III at $\alpha = 1$ only; for $1<\alpha<\alpha^*$ the sequence is I,II,I,III and for $\alpha > \alpha^*$ it is I,III. But these statements are specific to the model variant chosen.) For all of these regimes the ultimate state of the system at high stress is that of arrest.

I. This regime resembles full jamming (regime VI) except that $\sigma_{c2}$ is unbounded. The material is Newtonian at low stresses but above $\sigma_{c1}$ it is jammed into a solid, and remains so no matter how large the stress becomes. Note that there is a finite maximum shear rate for this flow curve.
 
II. This is a somewhat peculiar regime which resembles full jamming (regime IV) with upper and lower yield stresses $\sigma_{c1}$ and $\sigma_{c2}$. However, instead of showing shear thinning for $\sigma > \sigma_{c2}$ the material shear thickens again, and the flow curve makes a second collision with the vertical axis at $\sigma_{c3}$. For all stresses beyond $\sigma_{c3}$ it remains jammed, thus sharing the same ultimate behaviour as regime I, with a maximum flow rate.
For more details of this rather baroque regime, see Ref.\cite{jrheol}.

III. In this regime, with $v_2>4$ and $\alpha>1$, the material cannot flow by any steady homogeneous mechanism; the flow curve $\sigma(\dot\gamma)$ is simply a vertical line segment extending from the origin to infinity. In physical terms, the material is already a glass in its quiescent state at zero stress, and becomes ever more strongly arrested as stress is applied; it can never be shear melted. We call this the `no flow' regime.

\subsection{Trajectories in parameter space}
\label{threepointtwo}
Let us now consider how the control parameters $v_2,\alpha$ might translate into physically measurable quantities. For example, suppose we have hard-sphere colloids. At low concentrations $\phi$ these are not glasses, and not jammable either; we must have $\alpha <1$ and $v_2<4$. Clearly $v_2$ increases with $\phi$ and we can map the schematic model onto such colloids by demanding $v_2(\phi_g) = v_c = 4$. We also expect $\alpha(\phi)$ to be an increasing function -- dense colloids are more jammable -- but have no clear idea of this dependence. Depending on the form of $\alpha(\phi)$, the observed sequences on increasing $\phi$ could be IV,V,VI,VII; or IV,V,VI,VII,III. We assume here that $\alpha(\phi)$ is monotonic and that regime VII, the yielding glass, is mandatory (since it is observed in experiments); otherwise the list gets longer. However, if either condition is relaxed, or if a more general version of the schematic model is chosen, other regimes such as I, Ia, and VIa could also enter the sequence. Moreover if the interaction between particles is altered, for example by adding attractions \cite{pham}, still different regime sequences could arise including, for example, IV,I,III. 

\section{The physics of granulation}
\label{granulation}

Trajectories involving regimes I, Ia, and III are particularly interesting. In each case, the ultimate fate of the material is to be jammed at high stresses. (The same applies to the more baroque regimes II and IIa.) The simplest such case is regime III which, as described above, permits no steady flow at all. It seems plausible that this regime really exists in dense colloids, since it is known to exist in dry granular materials for $\phi > \phi_c$: if such materials are not allowed to expand, they will not flow homogeneously at any stress level. 
On increasing $\phi$ in a colloidal system, one can therefore expect a second threshold at some value $\phi_n$, beyond the usual colloidal glass transition $\phi_g$, marking a transition from the yielding glass (regime VII) into a non-yielding glass (regime III). For $\phi>\phi_n$, Brownian motion is not enough to bypass the need for dilatancy, which in turn is prevented by the fixed volume of solvent. Since Brownian motion is increasingly ineffective at large $\phi$, it seems plausible that $\phi_n$ is strictly less than $\phi_{rcp}$, and we assume this below. (However in principle these might coincide, in which case $\sigma_y$ could be expected to diverge smoothly at $\phi_n=\phi_{rcp}$.)

It seems clear that, if {\em sufficient} stress is applied, even the non-meltable glass of regime III must give way, somehow. This is likely to involve brittle fracture, plastic slip, or a related failure mechanism. In the early stages of granulation one may indeed observe something like brittle fracture, leading to creation of air-solvent interfaces. (The results reported in Section \ref{expt} are related to, but somewhat different from, this fracture picture: air is entrained in the form of large bubbles.) In other flow geometries the medium appears to dilate selectively in localized shear bands, allowing these to flow plastically \cite{haw,hawpc}; see \cite{cornell} for a related theory. (Note that this banding mechanism require an increase in $\phi$ through the remainder of the system -- which is increasingly difficult as $\phi_{rcp}$ is approached.)
 
For now, let us assume that regime III glasses are indeed brittle. In regime I, the stress-induced jammed state is brittle in the same sense. In regime Ia likewise, one has a shear-melting glass that later rejams into a brittle one. (Regimes II and IIa are also brittle at high $\sigma$.) If so, there is some stress $\sigma_b$ beyond which the vertical portion of the corresponding flow curves is effectively terminated by brittle fracture. 

It is hard to estimate $\sigma_b$ precisely but it surely involves the solvent-air interfacial tension $\Sigma$. Dimensionally $\sigma_b = \Sigma/\ell$ with $\ell$ some characteristic length. The smallest length in the problem is $a$, the particle radius; there could be a structural length scale (e.g., a force chain correlation length) that would be a multiple of this. But so long as we assume that the size of the sample does not matter, there is no other length scale than these, and hence $\sigma_b\simeq b\Sigma/a$ where $b$ is a (possibly small) prefactor. 
In comparison, the basic stress scale for a dense suspension of hard-sphere colloids is $k_BT/a^3$. This sets the vertical scale on flow curves such as those of Figure \ref{fulljam}, although the prefactor involved, like the elastic modulus $G$, should diverge at $\phi_{rcp}$ \cite{rcp}.
Accordingly we get a characteristic scale for these curves $\sigma_c\simeq G = c k_BT/a^3$ where $c$ is a prefactor that diverges at $\phi_{rcp}$. Within the glass phase ($v_2>4$) experiments give values in the range $10<c<1000$ as $\phi$ varies between 0.58 and 0.64 \cite{petekidis}, so values around 100 for $c$ are perhaps pertinent.

The two stress scales, $\sigma_b$ for fracture and $\sigma_c$ for homogeneous rheology, thus obey
\begin{equation}
\frac{\sigma_b}{\sigma_c} \simeq \frac{b}{c}\, \frac{\Sigma a^2}{k_BT}
\end{equation}
The ratio $b/c$ may be small, but the second factor is always very large; with (typically) $\Sigma = 0.1$ Nm$^{-1}$, $\Sigma a^2/k_BT$
is about $10^{7}(a/a_0)^2$ where $a_0 = 1 \mu$m, a typical colloidal size.
Therefore unless $a$ is extremely small or else $b=\ell/a$ is unexpectedly huge, there is always a good separation of scales between the fracture stress $\sigma_b$ and the stress scale $\sigma_c$ connected with the flow curve for homogeneous deformation. 

We may conclude from this that the physics of shear-melting of glasses, and/or jamming
into a stress induced glass, takes place on a much lower stress scale than
brittle fracture arising from ingress of air. Assuming no further physics intervenes at intermediate stresses, the picture of early-stage granulation that emerges is as follows. The quiescent fluid is arrested by jamming at some stress of order $\sigma_c$ (unless it is already a glass, which is arrested anyway). So long as we are in regime I, Ia or III, the material remains jammed until $\sigma>\sigma_b$, beyond which it breaks to bits by brittle fracture. Once broken, the pieces grind against each other in a complicated process which must somehow determine the final size of the granules, and which we do not attempt to model here, deferring instead to the extensive literature on this topic \cite{granulationmodels}. 

An important component of this picture is that the broken bits of the jammed state remain jammed. At first sight this might be paradoxical since the broken bits are only in loose mechanical contact with one another and the stress in a flowing state of the granules could be rather low -- lower than $\sigma_c$ even.
However, the granular state persists even when flow ceases, so we must anyway look elsewhere than the macroscopic rheological stress to explain the persistence of granules. The culprit is of course clear \cite{granulationrefs,granulationmodels}: capillary forces.

\section{The physics of granules}
\label{granules}

A similar separation of stress scales underlies the stability, including the remarkable bistability, of granules. With a flow curve like that of regime I (recall that this is the `maximum shear rate' version of full jamming, in which $\sigma_{c2}$ has diverged), a static jam can be maintained with any $\sigma > \sigma_{c1} \simeq \sigma_c$.

On the other hand, the capillary forces at the surface of the granule provide a stress of order $\Sigma/L$, where $L$ is the radius of mean curvature adopted by the fluid. Note that, viewed microscopically, $L$ is the same everywhere on those parts of the solvent surface not covered by particles. (This must be true unless the solvent {\em itself} ceases to be fluid; the static surface of such a fluid must have constant mean curvature.) However, $L$ could be much smaller than the radius $R$ of the granule itself. Moreover, given that the solvent fully wets the particles (which we shall assume for simplicity although other cases are possible), then once these protrude through the surface, the local curvature $L$ will have opposite sign to the macroscopic curvature determined by $R$. That is, the parts of the interface not covered by particles must be dimpled inwards to maintain a zero contact angle at the solvent-particle-air contact line. The maximum radius of curvature of such dimples is of order $a$; this sets the upper limit to the capillary stresses beyond which static equilibrium cannot be maintained. 

Exceeding this threshold would require a granular surface so dry that colloidal particles would be stranded there with almost no fluid around them (and would presumably be in danger of falling off to create some nearly dry powder). But in practice, for reasonable parameter choices, this upper limit to capillary stresses, of order $\Sigma/a$, far exceeds the stress of order $\sigma_c$ required to jam the interior of the granule. 

\subsection{Explanation of granulation phenomenology}

We are now in a position to offer explanations of the several phenomena concerning granulation listed in the Introduction.

\subsubsection{Granular stability:}
The preceding arguments show that there is no problem selfconsistently maintaining a jammed state of the granule in the manner outlined in the Introduction: capillary stresses at the granule surface sustain shear stresses within the jammed solid. A simple picture is to consider a set of linear force chains \cite{jamming} arranged like spokes of a wheel, each with its last particle poking slightly through the surface; each chain carries a compressional load caused by the interfacial tension which is equivalent to shear stresses at 45 degrees to the local (radial) force chain direction \cite{jamming}. Thinking about this in more detail, the force chain network required to support the radial normal stresses created by capillary forces must be more complex than just described, at least in three dimensions (if only to allow constant $L$ at the liquid surface).  But so long as there is such a network, the capillary forces can indeed create a compressive stress throughout the granule that is not isotropic, much larger than $\sigma_c$ in magnitude, and hence capable of sustaining the jammed state.

The above arguments apply throughout regimes I, Ia, II, IIa and III. (These are the regimes in which no homogeneous flow is possible at high stresses.) It is altogether more delicate to argue for self-consistently jammed granules in regime VI (or its variant VIa). Recall that this is the full jamming regime with re-entrant melting (Figure \ref{fulljam}). Here the the jammed state melts again on the stress scale $\sigma_{c2} \simeq \sigma_c$, and hence, in order to maintain jamming, the capillary-induced shear stress must {\em not} exceed this scale. If $\Sigma/R<\sigma_c$ \cite{footcond}, the required stress window looks easily achieved, but if not, some fine tuning of the surface structure of the granule is needed to sustain it: the microscopic liquid interface must be unnaturally flat on the scale of the macroscopic granular shape. We do not know whether such fine-tuning occurs in practice, nor, if it does, whether this imparts some special mechanical delicacy to the granules. Since the very weak curvature required anyway has opposite sign to the macroscopic curvature of the granule (as explained above), perhaps it makes no difference. In any case, this subtlety does not affect regimes I--III, in which there is no re-entrant melting of the homogeneous state at high stresses.

Note also a possible lower limit to the size of a jammed granule: if too small, the decrease in bulk volume fraction caused by particles protruding through the surface could mean that a jammed state is no longer sustainable in the interior. This decrease is of order $\delta\phi/\phi\sim \alpha a/R$ where $\alpha\le 1$ is the fractional volume of each surface particle that protrudes through the surface. This lower limit might be relevant to determining the granule sizes actually observed in granulation, but so long as it is exceeded, the above discussion is unchanged.

\subsubsection{Bistability:}
Even for a small granule with $\Sigma/R \gg \sigma_c$, there is no difficulty, in any of the relevant regimes, in explaining the bistability between the granule and the liquefied droplet state that was described in the Introduction. For, once the interior of a droplet is liquefied, its surface is spherical, and the capillary forces produce only an isotropic Laplace pressure in the interior. Although this may be formally large on the scale of $\sigma_c$, such a diagonal stress contribution is borne by the incompressibility of solvent and particles, not a contact network, and cannot jam the system. Thus the onset of granulation involves setting up a shear stress and/or a normal stress difference, not just a pressure; this anisotropic stress jams the bulk suspension, and promotes its brittle fracture. Subsequently capillary forces can maintain the local stress anisotropies that were set up initially within the granular interior, but they cannot create these from scratch. 

Clearly, bistability does require that the volume fraction of the colloid is low enough to sustain a flowable state at low stresses -- as happens for example in regime I of the phase diagram, Figure \ref{phasediag}. At extremely high colloid concentrations (for example in regime III) all states are jammed. Granulation in this limit is relatively trivial; a granule is maintained solid by capillary forces but would also be solid without them.
Similar physics is expected in fluidized bed granulation where a small amount of fluid is added to a dry powder \cite{grannote}; this will bind the particles into agglomerates stabilized by capillary bridges, but there is not generally enough fluid to create a spherical droplet of flowable particle concentration and bistability is presumably absent.

\subsubsection{Coalescence of Granule and Dropet:}
The coalescence of a flowable droplet and a granule is also explicable in this picture. Due to the inward curvature at the liquid surface of a granule (see above) the Laplace pressure in a granule is smaller than in a flowable droplet. Placing the two in contact will thus lead to a transfer of solvent and the collapse of capillary stress in the granule; it can then melt and coalesce with the adjacent droplet.

\subsubsection{Effect of Vibration:}
The conversion of a granule to a flowing droplet by vibration presumably involves melting of the contact network. This is consistent with the idea of fragility \cite{jamming} in which the network has a structure specifically adapted to the stresses applied. It is therefore quite possible that, even in an ultimately jamming regime such as I or III (as favoured above to explain the stability of a granule), a shear stress of order $\sigma_c$ applied as an ``incompatible stress'' will melt it. (An incompatible stress is a shear stress in an orientation other than the one the network has evolved to support \cite{jamming}; vibration would supply this.) Alternatively, in regime VI (and also VIa), any stress large compared with $\sigma_c$ will give the same effect. More complex explanations, involving disruption of the solvent/air interface by vibration, may also be possible.

\section{Concluding remarks}
The phenomenology of early-stage granulation in dense colloidal suspensions appears to be broadly explicable within a framework that combines capillary forces at the air-solvent interfaces with a schematic model of colloidal jamming. Several scenarios for this were developed above. Most of these involved a specific form of jamming in which simple shear-melting (of a glass) and/or re-entrant shear-melting (of a stress-induced solid) are excluded. In materials governed by such scenarios, sustained shear was argued to lead to brittle fracture, creating large amounts of air-solvent interface. The capillary forces generated at this interface are sufficient to maintain a jammed state of the granules which are then internally solidified by shear stresses. Bistability between a granule and a flowable droplet is then explicable, as is vibration-induced melting, and granule/droplet coalescence. The observation of such phenomena \cite{bibette,haw,warren,bibettepc} is perhaps some of the strongest evidence for jamming, in the specific sense of arrest caused by anisotropic stresses in dense colloids \cite{jamming,epl,jrheol}. The preliminary results of Section \ref{expt} suggest that this picture extends to well-characterized hard-sphere colloids as well as to systems with attractive interactions; however, these experiments are not yet conclusive and will be reported on more fully elsewhere \cite{hawpc}.

Since the details of the colloidal rheology depend strongly on concentration and on colloid-colloid interactions, it is possible that the favoured `brittle' form of jamming arises only in a limited parameter range. It would be very interesting to search for this directly in stress-controlled bulk rheology (which allows a jammed system to come to rest at finite stress). One could then see whether the onset of granulation under sustained shearing correlates with specific regimes of jamming as developed in the phase diagram of Figure \ref{phasediag}, and also whether the sequence of these regimes is in accord with the scenarios developed in Section \ref{threepointtwo}.

\ack

We thank Thomas Voigtmann and Patrick Warren for useful discussions.
This work was funded under EPSRC Grant GR/S10377/01.


\begin{thebibliography}{99}
\bibitem{grannote} In a second type of granulation, fluid is added slowly to a sheared dry powder, normally suspended in air as a fluidized bed. This is rather a different process which we do not address here, although once granules form, their agglomeration and breakup behaviour under continuing shear may be similar.
This paper concerns only
granulation by shear of a pre-existing dense suspension, sometimes called `high shear mixer granulation'.
\end{thebibliography}
\end{document}